\documentclass[fleqn,usenatbib,onecolumn]{mnras}
\usepackage{mathptmx}
\usepackage{gensymb}
\usepackage{graphicx}
\usepackage{caption}
\usepackage{hanging}
\usepackage{adjustbox}
\usepackage{bm}
\usepackage{amsmath}

\newcommand{\msun}{M$_{\odot}$}
\newcommand{\msunpc}{M$_{\odot}$ \rm{pc}$^{-3}$}
\newcommand{\Omm}{\Omega_{\mathrm{m}}}
\newcommand{\LCDM}{{\Lambda}{\mathrm{{CDM}}}}
\newcommand{\LCDMbf}{\pmb{{\Lambda}{\mathrm{{CDM}}}}}
\newcommand{\mum}{${\mu}\mathrm{m}$}

\begin{document}{\noindent\Huge\bf Rotation curves and the dark matter problem}

\vspace{0.5cm}

{\noindent\LARGE Albert Bosma - bosma@lam.fr}
\vspace{0.15cm}

{\noindent\LARGE Aix-Marseille Univ, CNRS, CNES, LAM, Marseille, France}

\LARGE
\section*{\LARGE Abstract in English}
The concept of dark matter in the Universe and its components has been
discussed in the 1930s by several authors, and in particular
by Oort (1932) and Zwicky (1933). However, it is only in the
1970s that the existence of dark matter
was considered convincing, thanks in part to
observations of the rotation curves of galaxies. This
dark matter should be present at multiple scales, in the
solar neighborhood, our Galaxy, near and distant galaxies,
clusters of galaxies, and the entire Universe. The subject attracts a
very large community to discover the nature of this material, without
achieving it. I will present my version of the history of this subject,
trying to shed light on some philosophical aspects.

\vspace{-0.5cm}

\section*{\LARGE R\'esum\'e en Fran\c cais}
Le concept de la mati\`ere noire dans l'Univers et ses composantes a \'et\'e 
discut\'e  dans les ann\'ees 1930s par plusieurs auteurs, et en particulier
par Oort (1932) et Zwicky (1933). Cependant, c'est seulement dans les
ann\'ees 1970s que l'existence de la mati\`ere noire 
a \'et\'e  jug\'e  convaincant, gr\^ace en partie
aux observations des courbes de rotation des galaxies. Cette
mati\`ere noire devrait \^etre pr\'esente \`a plusieurs \'echelles, dans le
voisinage solaire, notre Galaxie, les galaxies proches et lointaines,
les amas de galaxies, et l'Univers tout entier. Le sujet attire une
communaut\'e très large oeuvrant pour d\'ecouvrir la nature de cette 
mati\`ere, sans
y parvenir. Je pr\'esente ma version de l'histoire de ce
sujet, en
essayant d'\'eclairer quelques aspects philosophiques.


\section{\LARGE Introduction}
To understand the dark matter problem in astronomy, it is 
worthwhile
to start with some historical considerations. As Steven Weinberg remarks: "The best antidote to the philosophy of science is a knowledge of the history of science"
\citep{weinberg03}.
Although mention has been made of
``dark matter" before the 1930s (see, e.g.
\citealt{bertone18} 
for a review), 
the dark matter problem 
was considered on four different scales in four papers in the 1930s: 
in the solar neighbourhood 
\citep{oort32}, 
in the nearby spiral galaxy in Andromeda 
\citep{babcock39}, 
in the Coma cluster of galaxies 
\citep{zwicky33} 
and in a model of the Universe 
\citep{einstein32}. 
Hence this ``mass discrepancy problem" in galaxies and
cosmology has been
around in its entirety since the 1930s.

In this write-up of my talk, I will discuss each of these papers in turn,
and then move on to more recent developments, highlighting changes of 
insight, not only resulting from new data using better observing
techniques, but also from numerical simulations 
using increasingly more powerful computers.
My own involvement in this problem started in the 1970s, based on dark matter inferred from rotation curves, so I will discuss these in some detail. By the end of the 1970s, dark matter was widely accepted, and gradually incorporated in an overarching theory 
of cosmology and the formation and evolution of galaxies. Moreover, 
a connection to high energy physics was established, and the 
identification of this matter discussed. Now, more than
40 years later, dark matter has not yet been identified, although many proposals are under consideration. Some colleagues think the law of gravity should be 
modified, and I will discuss this briefly at the end. 

\section*{\LARGE 2. The 1930\lowercase{s} papers}

\subsection*{\LARGE 2.1 A model of the Universe}

A first discussion of the expansion of the Universe, based on Einstein's general relativity theory, was given in \cite{lemaitre27}, 
where he found from then available observations that the expansion
is linear, with a Hubble-Lema\^{i}tre constant of 625 km/s/Mpc. 
\cite{hubble29} 
presented a more accurate velocity - distance relation, calibrated partly with what he thought were bright stars, 
and found a similar value for the expansion constant. 
\cite{einstein32} 
considered a Universe
with enough matter so that it would expand for-ever, but at the critical
density separating a closed from an open Universe. 
Their ${\Omm = 1}$ model had a density close to the mass density 
of our Galaxy of 2.0 10$^{11}$ {\msun} in a box of 0.3 Mpc, as given by 
\cite{oort32}. 
This ${\Omm = 1}$ model was still preferred in 1985, at the first IAU Symposium on Dark Matter, by half of the voters sure of their opinion about this when asked in a discussion 
\citep{peebles87}. 

In an interesting article in the Dutch newspaper ``Algemeen Handelsblad",
\cite{desitter30} explains the model of an expanding Universe, and comments that the timescale for the expansion is of the same order as the lifetime of the Earth. He remarks that this is a less attractive aspect of the theory. 
This notion was later taken up by \cite{hoyle50}, who argued strongly for 
a steady state theory. However, \cite{sandage58} showed that the Hubble
constant was about 75 km/s/Mpc, a value much lower than those found in the 1930s, so that timescale problem disappeared.

\subsection*{\LARGE 2.2 The solar neighbourhood}
\cite{oort32} 
considered the vertical motions of stars in the solar neighbourhood, and found that the mean mass density should be
0.092 {\msunpc}. A careful examination of known stars resulted in a total
density of 0.038 {\msunpc}. The difference was thought to be fainter stars, meteors, and nebular material. In later literature, the
gap between the dynamical estimate and the stellar one was filled up, first
by the discovery of fainter stars, the presumed presence of stellar
remnants, 
but also of the presence of neutral atomic
hydrogen, as measured with the 21-cm HI line, and molecular 
hydrogen, as estimated from CO observations. A more modern 
study, by \cite{flynn06}, 
lists all these components, and the density of dark matter
in the solar neighbourhood was estimated at 0.008 {\msunpc}. 
A recent estimate based on Gaia data \citep{widmark21} 
arrives at a similar value.

\subsection*{\LARGE 2.3 Nearby galaxies}
Here I will mention a review
of \cite{lundmark30}, written in German, where he tabulated 
for 6 galaxies the ratio of
bright + dark matter / bright matter, and found it to vary between
6:1 (for the galaxy M33), 10:1 (for the Milky Way and M51), 20:1 (M31), 30:1 (NGC 4594) and 100:1 (M81). This work, not often referred to, was noted in a 
talk by \cite{bergstrom15}, and made its way into a book
review by \cite{trimble21}. However, a
clear exposition of this work in English is available in
\cite{lundmark28}. He makes a distinction between the 
gravitational mass, derived from the spectroscopic 
observations of rotation, and the
luminous mass, derived from the brightness of a galaxy and an 
assumed mass-to-light ratio. The 
gravitational mass will also include the dark matter due to 
extinct stars, dark nebular matter and meteors.
For M81 he used data based on a spectrum by Max
Wolf, but the scatter is very large when compared with
modern data by \cite{vegabeltran01}. 
The value for the luminosity of M81 is different
in Lundmark's two papers, which is very confusing.

\subsection*{\LARGE 2.4 The Coma cluster of galaxies}
\cite{zwicky33},
in a paper written in German, came to an interesting conclusion by considering the apparent motions of 8 galaxies in the Coma cluster, based on work by 
\cite{hubble31}.
The apparent velocity
dispersion he found was about 1000 km/s. By applying the virial theorem,
he derived an expected velocity dispersion of 80 km/s, assuming the cluster
to contain 800 galaxies of 10$^9${\msun} each. He attributed the discrepancy
to "dunkle Materie", i.e. dark matter. The density he found was roughly 
comparable to the one in the model of \cite{einstein32},
as remarked in a footnote. For the Virgo cluster, 
\cite{smith36} arrived at a similar conclusion. Either
the nebulae in Virgo had a mass of 2 
10$^{11}${\msun}, i.e; much larger than the value found by 
\cite{hubble34}, or there is internebular material, in
agreement with Zwicky's result for the Coma cluster. 

The mass of the Coma cluster was reconsidered in \cite{schwarzschild54}, in 
a wide ranging paper discussing also the masses of galaxies at smaller scales. Schwarzschild used the velocities of 22 galaxies in the cluster, made available to him by Humason. 
He found a velocity dispersion of
1080 km/s and derived a typical mass-to-light ratio per galaxy of 800 in solar units. He concludes that
"this bewilderingly high value for the mass-luminosity must be considered as very uncertain, since the mass and
particularly the luminosity of the Coma cluster are still 
poorly determined." 

References of either 
\cite{smith36} or \cite{schwarzschild54} 
are not listed in the NASA/ADS system, which led to a modern 
myth that Zwicky's work did not get the attention it 
deserved in those times. Internebular material was 
found much later, by interpreting the X-ray radiation as coming from hot 
intergalactic gas in clusters of galaxies
(e.g. \citealt{felten66}, 
see also \citealt{sarazin86} 
for a review), as exemplified by 
modern data for the Coma cluster in \cite{churazov21}.

\subsection*{\LARGE 2.5 The rotation curve of the Andromeda galaxy}
\cite{babcock39} 
determined the rotation curve of M31, the Andromeda
galaxy, from spectra obtained with the 36-inch telescope at 
Lick Observatory. He used absorption lines to determine 
velocities in the inner
parts of the galaxy, and emission lines of a few bright 
nebulae in the outer parts, some of which were observed for 3 
nights in a row to get $\sim$20 hours of exposure time. 
He found a rising rotation curve,
and argued for dark material (also in 
\citealt{babcock47}), 
but his data have not been confirmed by more modern work.

Later work on M31 by Mayall and Adams was presented in \cite{schwarzschild54}, 
who could fit a mass model with a constant mass-to-luminosity ratio to the data, out to the largest radius of 
$\sim$120$^\prime$. \cite{vandehulst57} 
observed M31 in the 21-cm HI line with the Dwingeloo 25-m radio telescope, and derived a rotation curve out to $\sim$150$^\prime$. Analysis of this curve by 
\cite{schmidt57} 
showed that the mass-to-luminosity ratio could
possibly increase as function of radius in the outer parts. This prompted \cite{devaucouleurs58}  
to do photoelectric photometry of this galaxy. His 
determination
of the mass-to-luminosity ratios shows that they could
increase with radius, although tentatively.

\section*{\LARGE 3. Late 1950{\lowercase{s}} and 1960{\lowercase{s}}}

\subsection*{\LARGE 3.1 Mass-to-luminosity studies of galaxies, groups and clusters}

In a series of papers published in 1959 - 1965, Burbidge, Burbidge and Prendergast, later on joined by Rubin and Crampin, presented rotation curves using the emission lines of H$\alpha$ and [NII] of $\sim$30 galaxies. These curves extend typically out to about
one half of the optical radius. Their modelling procedure is given in
\cite{burbidge59}, 
and their reputedly `best' curve is  the one for NGC 5055 
\citep{burbidge60}. 

\cite{kahn59} 
considered the situation in the Local Group, where
the main galaxies, M31 and the Milky Way, deviate from the Hubble flow,
and are approaching each other. They derive from this that the total mass
in the Local Group must be at least 1.8 10$^{12}$ {\msun}, and suggest
that most of the mass is in intergalactic gas. They ascribe the warp of the
HI layer of the Milky Way to the movement of our Galaxy through this gas.

\cite{page52, page65, page67}
studied binary galaxies and groups. He found that
spiral-spiral pairs have a low mass-to-luminosity ratio, while pairs of
early type galaxies (ellipticals or S0-galaxies) this ratio is much higher. In Figure 1, he shows that the mass-to-light ratio increases as function of
the number of galaxies in pairs, groups and clusters. 

\begin{figure}
\hspace{10pt}
  \begin{minipage}[c]{0.45\textwidth}
    \includegraphics[width=\textwidth]{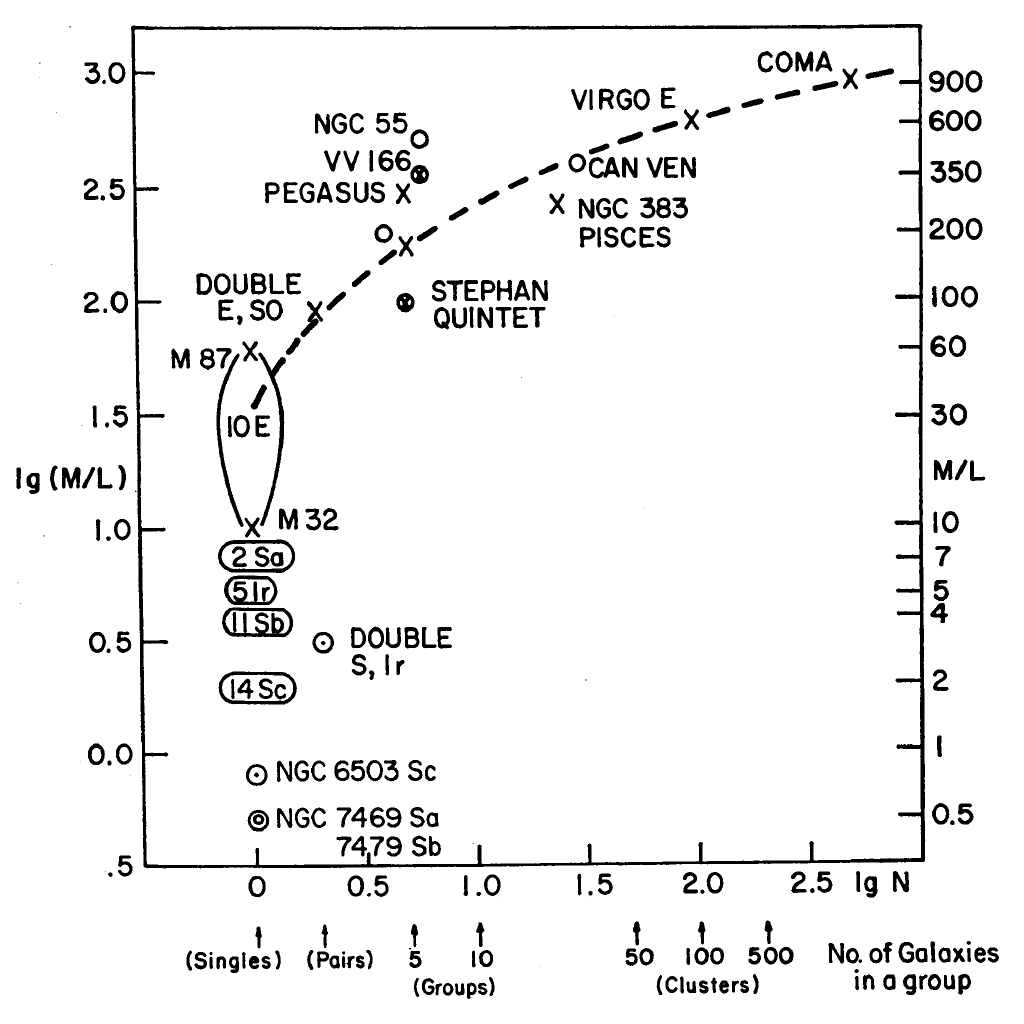}
  \end{minipage}\hfill
  \hspace{-10pt}
  \begin{minipage}[b]{0.5\textwidth}
    \caption{Average values of M/L for galaxies, using a Hubble Constant of
 100 km/s/Mpc. The data below M/L = 10 is for galaxies observed by Burbidge
 et al. 1959 - 1965, in part averaged by Hubble type Sa, Sb or Sc.
 These data do not indicate the presence of dark matter in these galaxies. 
 From \protect\cite{page67}.}
\label{fig:03}
  \end{minipage}
\end{figure}

\subsection*{\LARGE 3.2 Cosmology and the origin of the elements}

The cosmological debate in the late 1940s started with suggestions by \cite{gamov46} 
and \cite{alphabg48}
about the nucleosynthesis
of elements early on in the Universe. Analysis of these ideas led to a prediction for the temperature of the
Universe at the present time to be 5°K \citep{alpher48b}, and for the distribution of element abundances as function of atomic weight, which compared well with the then available observations \citep{alpher48a}. 
Further information about this problem can be found in a 
collection of essays in \cite{peebles09}, in particular
in a diagram in its section 3.1, as well as in a recent book by \cite{peebles20}.


A different theory was developed by \cite{hoyle48},
and \cite{bondi48}, and became known as the ``steady state" theory. \cite{hoyle50}
discussed his objections to the ``Big Bang" idea at the time, noting that the predicted value of 5$^\circ$K for the microwave background was above the temperature
of the interstellar medium as discussed in 
\cite{mckellar41}, 
and that the time since the expansion began is less that the accepted lifetime of present constituents of the Universe (see also section 2.1). This turned out to
be a ``naive" falsification, since Hoyle tacitly assumed that the then current values of both the predicted background temperature and the Hubble constant were correct. Warnings about using
such a simplified philosophical approach have been
given by \cite{duhem06, duhem91}. 

Indeed, one of the advances in the 1950s was the revision of the Hubble constant. Red sensitive photographic
plates became available, enabling \cite{sandage58} 
to re-address the distance scale in the Universe using Cepheids, and finding a value of H$_0$ of 75 km/s/Mpc. Another advance was the extensive study of 
\cite{burbidge57} on the formation of the elements in stars.

Present day studies still address the H$_0$ ``tension", between the values derived on cosmological scales compared to those based on Cepheids in the local Universe (e.g. \citealt{divalentino21}), 
even though the discrepancy
is much smaller. 
Likewise, there are still some ``problems" with the picture of primordial nucleosynthesis 
(cf. \citealt{coc14}).

\vspace{-0.8cm}

\section*{\LARGE 4. Rotation curves of nearby spiral galaxies: the 1970{\lowercase{s}}}

\subsection*{\LARGE 4.1 Early work}
\cite{freeman70} 
studied photometric radial profiles of 36 disk galaxies, for which surface photometry in the blue wavelength band was available. He shows that the radial profile can be described rather well with an exponential disk, and finds that 
for 28 out of 36 galaxies the (extrapolated) central 
surface density of this disk is almost constant. 
He also derives a formula for its rotation curve, and
compares the prediction of such a curve for 
four galaxies with sufficient data. The Large and Small Magellanic Clouds
follow more or less the prediction, while the other 
two, NGC 300 and M33, do not. In these galaxies
``there must be additional matter which is undetected, either
optically or at 21 cm. Its mass must be at least as large as the mass
of the detected galaxy, and its distribution must be quite different 
from the exponential distribution which holds for the optical galaxy." 

\cite{rubin70} 
published an optical rotation curve of M31, based on spectra 
of 67 bright HII-regions, using 
the DTM image-tube spectrograph. Their data compare well with 
single dish 
21-cm HI data from 
\cite{burke64}
at 10$^\prime$ resolution. Later HI data by 
\cite{roberts75} 
show that the rotation curve remains flat out to radii 
beyond the optical radius. A comparison of several rotation curves for M31, and for a model exponential disk 
are given in Figure 2a. The HI data of 
\cite{vandehulst57}
and certainly those of 
\cite{roberts75}
go out 
far enough in radius to demonstrate the mass-discrepancy between the
expected and observed rotation curves, but the 
\cite{rubin70} 
data do not.

\begin{figure*}
\begin{adjustbox}{width=\linewidth}
    \includegraphics[height=3cm]{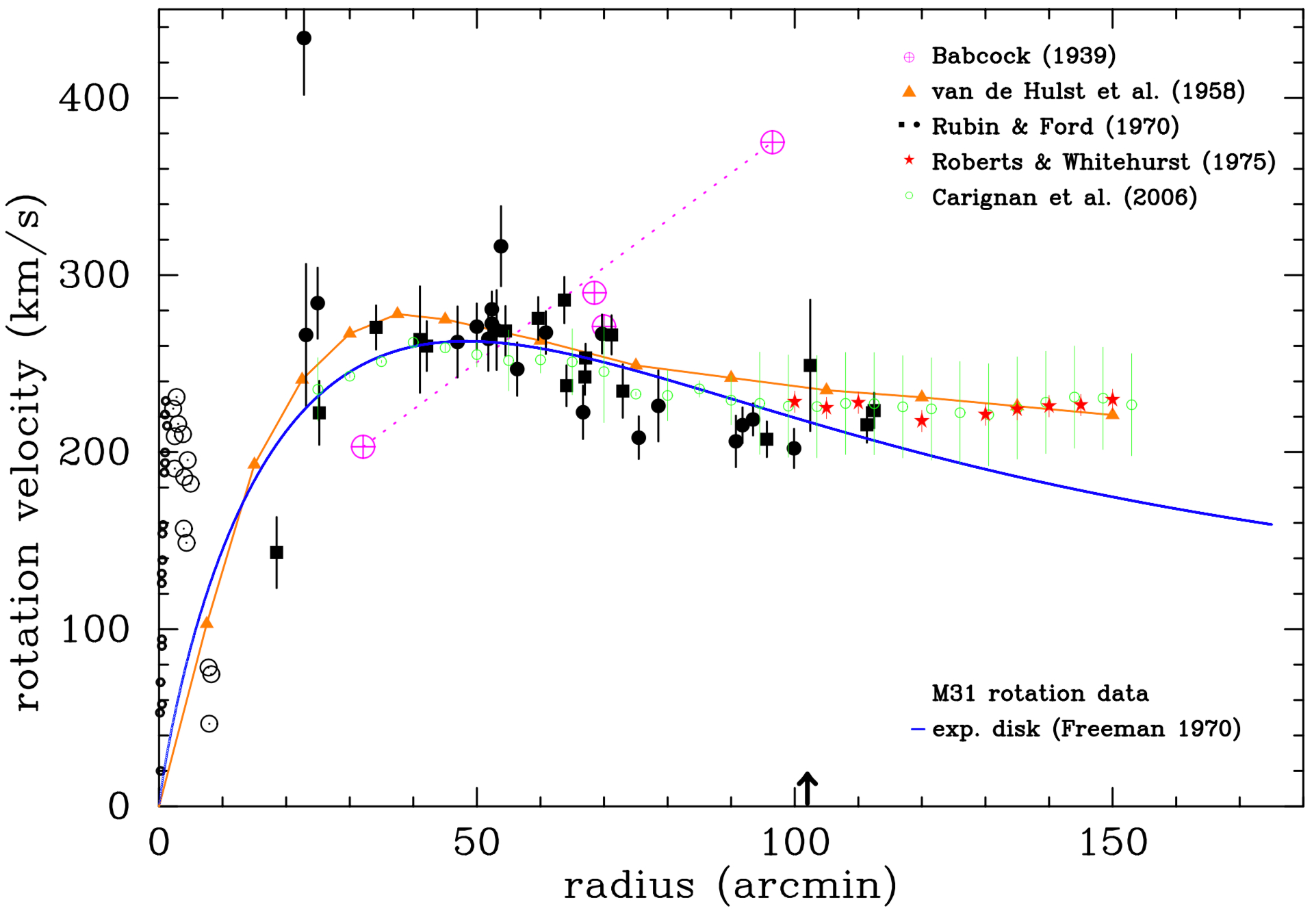}
    \includegraphics[height=3cm]{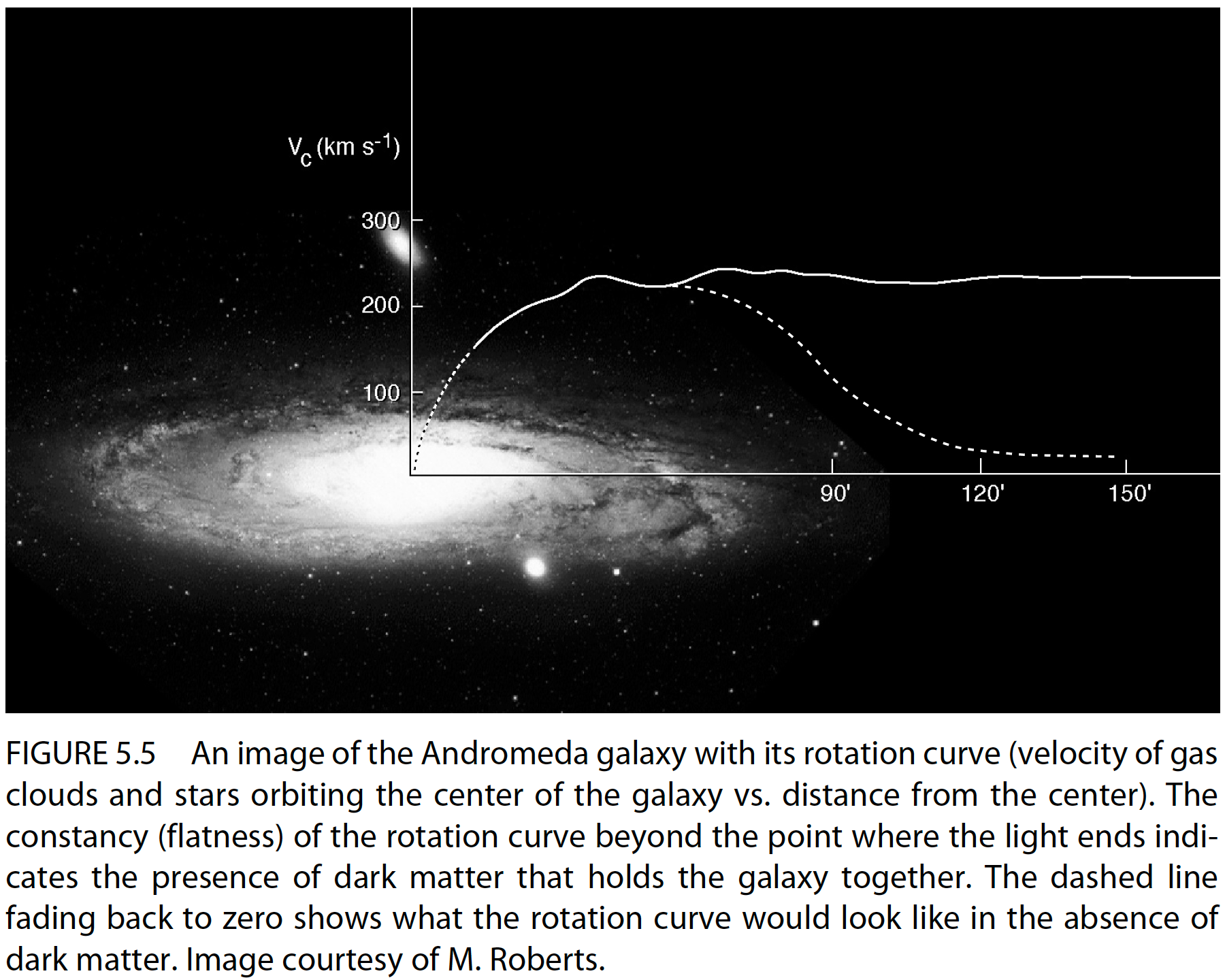}
\end{adjustbox}
\caption{a. At left: Rotation curves of M31, as determined by 
 \protect\citealt{babcock39}, (purple points), 
 \protect\citealt{vandehulst57}, (orange points), 
 \protect\citealt{rubin70}, (black points), 
 \protect\cite{roberts75},  (red points), and 
 \protect\cite{carignan06},  (green points). The blue line is a "maximum disk" rotation curve of an exponential disk with the scale length given in \protect\cite{freeman70}, 
 based on the study of \protect\cite{devaucouleurs58}. 
 The arrow indicates the optical radius of {102$^\prime$}.
 b. At right: Example of a predicted "Keplerian decline" of the rotation curve of M31,  \protect\citep{turner03}, 
 variants of which are still shown now in some talks. This seems due to the comparison of the
rotation curves of galaxies with the one
expected for the solar system: in the latter most of the mass is centrally
concentrated (for 96\% of the total mass) in the Sun, while for an exponential disk the distribution follows a much more gentle radial decline. Even so, a Keplerian decline means that at 3 times the radius where the rotation curve peaks, the velocity has dropped by a factor 1.73, to a value not close to zero.}
\end{figure*}

\subsection*{\LARGE 4.2 Misunderstandings}
It is of interest to compare Figure 2a with ``popular" graphs 
of the mass discrepancy in M31 shown mainly in non-
astronomical circles now participating in the study of dark 
matter, which show unfortunately 
a clear misunderstanding of the problem. 
There are several problems with Figure 2b: The real data, as 
given by \cite{roberts08},
extend to 150 arcmin, and not to 170 arcmin. The 
expected curve is very unrealistic, since for a Keplerian rotation curve 
the rotation varies with radius as R$^{-0.5}$, i.e.
at 3 times the radius of the peak value of the rotation curve, the value is 1/1.73 lower. Moreover, Figure 2a shows that the data of \cite{rubin70} are inconclusive.

Yet this work is frequently attributed to Rubin, instead of to Roberts. 
A further mystery is occasionally
created by the attribution to Rubin of later, more
convincing, data about galaxies such as NGC 3198, discussed in \cite{vanalbada85}, 
or NGC 6503, discussed in \cite{begeman91}.  
To quote from \cite{freese16}: 
"The tendency is to depersonalize
scientific discoveries and in the end to attribute them to one person.
But in reality, it is the collaboration of many people with sympathetic
scientific views and outlooks that solves problems - and the hunt for
dark matter is no exception".

\subsection*{\LARGE 4.3 The disk stability problem}

When computing power became sufficiently available in the early 1970s, 
several authors began to simulate disk galaxies. As initial condition,
the disk was represented by particles following the exponential disk
model, and given a certain initial velocity dispersion, enough to keep
it stable to axisymmetric instabilities: a Q factor larger than 1.0 
\citep{toomre64}.
The most striking result was that the disk formed a bar
\citep{miller70, hohl71}. 

To cure this problem, a radical solution was proposed in 
\cite{ostriker73}. 
They embedded the disk in a spherical halo potential,
and found that such a configuration remains more stable. They reasoned
that for our Galaxy, this requires a halo mass interior to the disk 
which is about equal to the disk mass. This could mean that the halo
mass exterior to the disk may be extremely large.
As a follow-up, \cite{ostriker74} 
considered a number of mass indicators, and found a near linear
increase of the mass in galactic systems as function of radius. A 
similar argument was made by 
\cite{einasto74}. 
For the mean density in the Universe,
Ostriker et al. argued that $\Omega$ = $\rho$/$\rho_{crit}$ $\sim$ 0.2,
with an uncertainty of at least a factor of 3.

In the early 2000s, the increased computing power enabled
\cite{athanassoula02, athanassoula03} to find
that the use of a live halo, i.e. one
composed of particles which could individually respond to forces, makes
a lot of difference compared to a fixed potential: angular momentum 
exchange becomes possible, and reinforces the bar. This result stands 
when gas is added to the simulations, cf. 
\cite{athanassoula13}. 
   
Debate followed for several years about the 1973/4 proposals and results.
\cite{burbidge75} 
provided strong criticisms. He pointed out that for M31
the increase in mass was relatively modest, and that at large radii the
mass indicators became very uncertain. He also argued that 
\cite{einasto74}
had rejected the presence of non-circular motions in the outer parts of
spirals without adequate justification. Others, e.g. 
\cite{materne76} 
argued that there are membership problems to consider for groups
of galaxies. Moreover, for pairs of galaxies, several new projects were
set up.

\subsection*{\LARGE 4.4 Radioastronomy contributions}

In the late 1960s, early 1970s, several improvements were realized in
radioastronomy, which led to the use of interferometers to do observations
of nearby galaxies in the 21-cm HI line. With a single dish, such as the 300
foot transit telescope at NRAO, Green Bank, the angular resolution was
about 10$^\prime$, which is useful for M31, but not for many other galaxies. With the two-element Caltech interferometer, angular resolutions
of 2$^\prime$-4$^\prime$ became possible, and results for HI rotation
curves for sample of 5 Sc galaxies were presented by 
\cite{rogstad72}. 
They modeled their galaxies
with Brandt curves, after 
\cite{brandt60}. 
However, the behaviour of such 
curves is only approximate: for NGC 2403, Figure 4 in 
\cite{shostak73} 
shows a slowly increasing rotation curve, while the fitted Brandt curve in 
\cite{rogstad72} 
slowly decreases.

Further improvement in angular resolution was obtained by the use of the
Westerbork Synthesis Radio Telescope (WSRT). In its initial phase, it 
consisted of 10 fixed and 2 movable antennas; later 2 other movable
antennas were added. In the 1970s, most results had an angular resolution 
of 25$^\prime$$^\prime$-50$^\prime$$^\prime$. 
\cite{roberts73} 
presented early data on 4 galaxies:
our Galaxy (using the curve in \citealt{schmidt65}), M31 (using 
data obtained with the NRAO 300 foot dish), M81 (using early
WSRT results from \citealt{rots75}) and M101 (
Caltech interferometer results from 
\citealt{rogstad71}). 
In retrospect, these data are
not very convincing, since M81 is the main galaxy in a group with M82 and
NGC 3077, and the kinematics of the HI envelope is rather asymmetric. The
rotation curve declines with radius in the SE part, while it increases
with radius in the NW part. M101 is very asymmetric in the outer parts,
as later shown clearly in WSRT data by 
\cite{bosma81c}. 
For our Galaxy, the situation was not very clear in the 1970s, so that 
the main result of \cite{roberts73}
rests in fact on the behaviour of the rotation curve of M31.

In \cite{bosma78, bosma81a, bosma81b} 
I presented rotation curves for 25 galaxies, by
compiling data from my own work with the WSRT, work of others with the WSRT,
and older data available already, with the criterion that the ratio of 
radius/beamsize was sufficient to obtain at least 5 - 10 independent points
on the rotation curves. Where possible, optical data were added in the inner
parts, so as to avoid beamsmearing effects.

The most interesting results I obtained in my thesis 
work, apart from the rotation curves, concerned the 
discovery of very extended HI distributions of some 
galaxies. A good example is NGC 2841, presented in
Figure 3, and combined
with a composite bulge-disk-halo mass model derived much
later, using data from the S4G survey (see 
\citealt{bosma17} for a more detailed description).

\begin{figure*}
\centering
\includegraphics[width=1.0\textwidth]{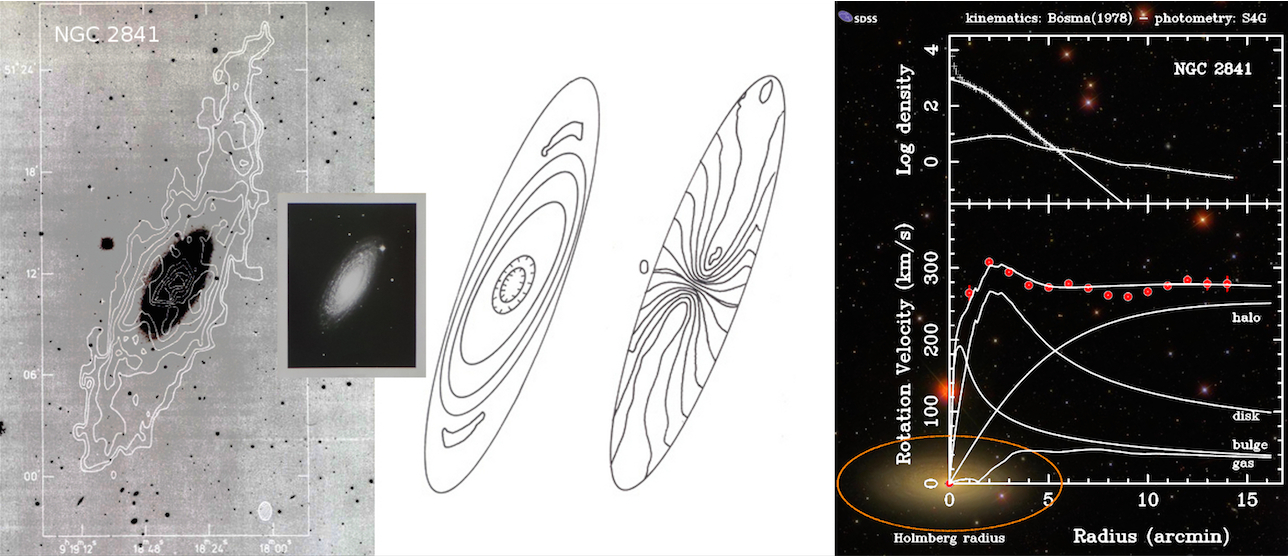}
\caption{At {\it left}, the H{\sc i} distribution in the galaxy NGC~2841 observed by \citet{bosma78}, overlaid on 
a deep IIIaJ image provided by H.C. Arp; the inset shows the Hubble Atlas image (\citealt{sandage61}).
The {\it middle} panels show the warp model. At {\it right} a mass model
of the galaxy adjusted to the H{\sc i} rotation data in \citet{bosma78},
calculated as described in \citet{athanassoula87}, using surface photometry data from the
 {\it Spitzer} Survey of Stellar Structure in Galaxies (S$^4$G; \citealt{munoz15}), and overlaid on scale on a Sloan Digital Sky Survey  (SDSS) colour image. The orange ellipse 
around the galaxy outlines the \citet{holmberg58} dimensions. Montage as in \citet{roberts88}, 
suggested in this form by Bernard Jones (private communication).}
\index{NGC~2841}
\index{warps}
\label{fig:n2841}
\end{figure*}

\subsection*{\LARGE 4.5 Summing up}

\cite{faber79} 
summed up the dark matter debate in the 1970s
by reviewing all the evidence for it, starting with rotation curve data, and reproducing my plot of the rotation curves of 25 galaxies. 
They also discussed in detail other evidence for dark matter, at larger
scales, i.e. pairs, groups and clusters of galaxies. 
Their review was quite influential: not only astronomers paid 
attention, but also particle physicists.
In a retrospect, \cite{rubin03} 
cites my plots and her visual 1978 spectra
as the most convincing evidence for the flatness of galaxy rotation curves,
and ``therefore" (her opinion) of dark matter in galaxies.

\subsection*{\LARGE 4.6 The radial extent of rotation curves matters}

While it is true that flat rotation curves can indicate the presence of
dark matter, this is not so straightforward if the extent of the curve is
limited. This was demonstrated in 1982 by 
\cite{kalnajs83}, 
in memorable oral contributions in meetings in Besan\c con and Patras (for more details on his approach see \citealt{kalnajs99}). 
What is occurring here was dubbed the 'bulge-disk-halo conspiracy' in 
\cite{bahcall85}. 
Somehow, the contributions of the bulge and the disk, using suitable 
mass-to-light ratios, can be found to fit the inner parts of a rotation
curve. If the rotation curve extends to a sufficient extent in radius, the
possible contribution of a dark halo may become necessary.

\cite{kent86} 
followed up on this, using models based on his CCD photometry
observations of a large number of galaxies in the samples of 
\cite{rubin78, rubin80, rubin82, rubin85}.
He found that most of the galaxies studied by Rubin, Ford et al. did not show
evidence for dark matter based on his rotation curve fits, while galaxies
based on HI data did (cf. \citealt{kent87}).
\cite{rubin89} remarked: ``Ten
years ago \cite{bosma78} wisely produced rotation curves by
joining the inner parts determined optically with the outer
parts determined from 21 cm observations. Such composite
curves remain the closest approximations to accurate 
rotation curves for spiral galaxies."

\subsection*{\LARGE 4.7 The disk-halo degeneracy}

The disk-halo degeneracy describes the situation that there is no unique
mass-to-light ratio for the disk: one can obtain fits to the 
rotation curve with a multitude of models involving the 
combination of massive disks and
light dark halos, while the opposite is also true. This was 
discussed in a graphical form by \cite{vanalbada85}. 
So how to constrain this further?
Intuitively, a solution of 'maximum
disk' seems more pleasing, but this is by no means certain. 
\cite{athanassoula87} introduced criteria based on spiral
structure theory: it was argued that a galaxy disk should 
be massive enough so that a spiral wave could develop. Since most spirals are 2-armed and relatively symmetric, 
the presence of an m = 2 spiral wave should
be required, while the presence of an m = 1 component should be
suppressed. This is illustrated in Figure 4b for the galaxy NGC 3198.

Later on, more criteria were applied to investigate this 
problem, such as considerations about our Galaxy, gas flows
in barred spirals, and spiral structure calculations.
A particular effort was made in the DiskMass project, 
using the behaviour of the velocity dispersion
in stellar disks (\citealt{bershady11, martinsson13}). The 
main result from this survey indicates
that disks are NOT maximum, in particular for lower mass
galaxies, which contradicts the earlier results. This is
shown in Figure 4a, where the caption provide further details.
Further studies
by \cite{aniyan18, aniyan21} tried to confirm this, though without really
convincing results. It is hoped that a new survey, with the
WEAVE spectrograph \citep{jin22}, will shed more light on
this question.

\begin{figure}
\includegraphics[width=1.0\textwidth]{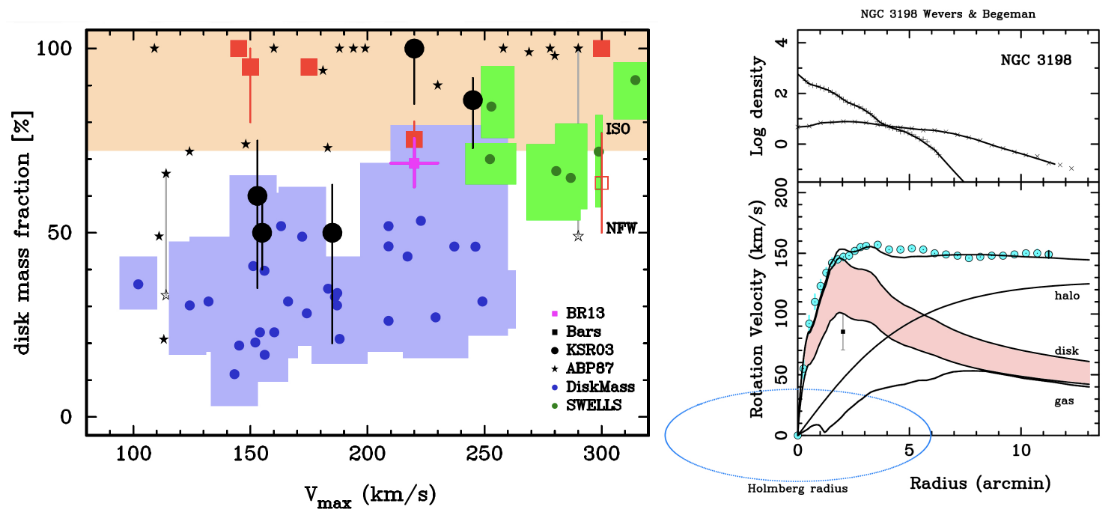}
\caption{
At left: disk mass fraction as function of the maximum velocity of the rotation curve,
determined with several methods. “BR13” \protect\cite{bovy13}
indicates the value for our Galaxy.
“Bars” concern the determination using gas flow models in barred spirals (see text). “KSR03”
concern five galaxies studied by \protect\cite{kranz03}. Black filled stars concern the results from 
\protect\cite{athanassoula87} for their “maximum disk with no m = 1” models, except for two thin vertical
lines at Vmax = 114.0 and 280.0 km/s which indicate also the “no m = 2” models. For the DiskMass
project, the results are taken as in \protect\cite{courteau15}, but the error bars are replaced by the
area spanned by them. A similar representation has been done for the SWELLS survey (\protect\citealt{barnabe11, dutton13}). At right: mass models for NGC 3198 using the method
described in \protect\cite{athanassoula87}, shown in detail in \protect\cite{bosma99}.}
\label{disk-halo-problem}
\end{figure}

\vspace{-0.4cm}
\section*{\LARGE 5. Wider dark matter problems}
\vspace{0.2cm}

\subsection*{\LARGE 5.1 Neutrinos and the dark matter problem}
Neutrinos were suggested as a candidate solution to the dark matter
problem as early as \cite{gershtein66}, \cite{cowsik72} and \cite{lee77}. This is due to the existence of the solar neutrino problem (cf. \citealt{cleveland98}), which probes the reaction 
$4{\rm{p}}  \to {}^4{\rm{He}} + 2{e^ + } + 2{\nu _e} + 26.73{\mbox{ MeV}}$. The reaction ${\nu _e} + {}^{37}{\rm{Cl}} \to {}^{37}{\rm{Ar}} + {e^ - }$, suggested by \cite{pontecorvo46}, allows detection of solar neutrinos with a large tank of tetrachloroethylene (CCl$_4$) in a shielded environment. The mismatch between the detected ones and the expectations led to the theory of neutrino oscillations, where electron neutrinos change flavour into muon and tau neutrinos, implying that neutrinos have mass.

A bold claim that neutrinos could be the "missing mass" in the Universe followed the announcement of \cite{lubimov80} of the detection of a non-zero electron antineutrino mass, but this was never confirmed. A newer candidate is the sterile neutrino
\citep{dodelson94}. The theory of "MOdified
Newtonian Dynamics" (MOND), developed by \cite{milgrom83a, milgrom83b} and discussed later in section 8, can be rescued in the case of galaxy clusters by 
postulating the existence of 11eV sterile neutrinos \citep{angus09}.
The latest experiment, MicroBooNE, found no evidence for 
light sterile neutrino oscillations \citep{microboone22}.
Analysis of the Cosmic Microwave Background by the \cite{planck13, planck18} gave upper limits for the neutrino background flux of 0.23 eV, and 0.12 eV, respectively. 

\vspace{-0.2cm}
\subsection*{\LARGE 5.2 Limits on neutrino mass from astronomical considerations}

\cite{tremaine79} argued that, based on phase space arguments, there is a lower limit to the mass of neutrinos which can be derived from the properties of e.g. clusters of galaxies. This can be used in to derive lower limits for the neutrino mass for dwarf spheroidal galaxies. In particular, the observations by \cite{aaronson83} of the Draco dwarf galaxy, in which he only observed 3 carbon stars, imply a lower limit of 500 eV (see also \citealt{lin83}). This exceeds the mean density of the Universe, of order 50 - 100 eV, so that neutrinos cannot be the dark matter in dwarf spheroidals.

Later on, cosmological numerical simulations advanced well enough to predict the character of the large scale structure of the galaxy distribution. The results of these simulations exclude neutrinos as a viable dark matter candidate (\citealt{white83}, see also Figure 2b in \citealt{davis92}).

\vspace{-0.2cm}
\subsection*{\LARGE 5.3 Gravitational lensing, concordance model}

An interesting system of gravitational lensed images is the 
so-called
Einstein Cross, Q2237+0305, discovered by \cite{huchra85}, 
where there are 4 images of the nucleus due to lensing. 
\cite{trott10} studied
the mass distribution of this galaxy, by combining rotation
curve data and lensing data, and argued that the
stellar bulge dominates the central parts of the mass 
distribution, rather than the dark matter.

In clusters of galaxies, the current practice is to
investigate the presence of dark matter using 
gravitational lensing. A first result was obtained by
\cite{soucail87,soucail88} for an arc feature in the cluster Abell 370. These investigations have now become a standard
approach to the dark matter problem in clusters of galaxies.
A particularly intriguing study was done for the Bullet
Cluster \citep{clowe06}, which is a system of two clusters 
in collision,
where the gravitational mass according to the lensing is
separated from the behaviour of the X-ray gas. It is
argued that this separation presents difficulties for a
theory based on modification of the gravity law.

In the 1990s, there was a lot of debate about the question 
whether the mass density of the Universe was close to the 
critical density (the $\Omega_{\mathrm{m}} = 1$ model). A summary paper
by \cite{ostriker95} proposed that the Universe has a low 
mass density, and that there is a cosmological constant: their
best model had a Hubble constant of 70 km/s/Mpc, 
$\Omega_{\mathrm{m}} = 0.35$ and
$\Omega_{\mathrm{\Lambda}} = 0.65$. Later on, thanks to
observations of supernovae (\citealt{perlmutter98,riess98}), the $\Omega_{\mathrm{\Lambda}}$ term was baptized dark energy.
The cosmic microwave background data
\citep{planck18} indicate a Hubble constant of 67.7 km/s/Mpc.

\vspace{-0.4cm}
\section*{\LARGE 6. What could the dark matter be?}
\vspace{0.2cm}

The short answer is: we do not know (yet), so I will limit the discussion here to a couple of remarks.

\subsection*{\LARGE 6.1 some philosophy}

A first comment is that the meaning of the word ``dark matter" has changed in the course of time. In the 1930s, it was meant to be
faint or extinct stars, dark nebular material and meteors, but in the 1980s it changed towards non-baryonic material, related to high energy physics. So the ostensive definition of dark matter has changed.

A second comment concerns the influx of physicists into astronomy,
which led in some places to a fusion between the
departments of physics and astronomy. Kevin Prendergast, who
worked with the Burbidges on their rotation curve project discussed
in section 3.1, articulated once the methodological distinction between physics and astronomy: ``Anything big enough to see is unique. Then what is important, or unimportant?
Physicists don’t face this question: all protons are the same, and everything about
them is important. Always seek the simplest situation in which the phenomenon
appears---we don’t have that luxury" 
(\citealt{prendergast13}).
In any case, the fact that nothing came after $\sim$40 years
of very optimistic stories about the detection of WIMPs (Weakly Interacting Massive Particles), etc., has led some people to question the whole procedure of 
doing science this way. No wonder that searches for an alternative, be it MOND or something else, are being
considered.

\subsection*{\LARGE 6.2 direct detection}

When the Large Hadron Collider started working, it was
expected that new particles would be easily detected, 
which could give clues to a possible dark matter candidate. 
However, except for the discovery of the Higgs boson, no new particle has been found yet. 
A comparison of the relevant figures in \cite{bertone10}
and \cite{bagnaschi17} shows that the theoretical 
predictions for direct detection change as the years move on.
Bertone pointed out in 2010 that ``the moment of truth has 
come for WIMPs: either we will discover them in the next 
five to ten years, or we will witness their inevitable 
decline." A description of the latest efforts which are
going to be made is in
\cite{aalbers23}.
Soon the astrophysical neutrino ‘floor’, below which 
astrophysical neutrino backgrounds dominate, will be reached, 
without any detection.

 \subsection*{\LARGE 6.3 indirect detection}

 There is a discussion about a possible indirect detection
 of dark matter, due to observations of the Galactic Center
 at $\gamma$-ray wavelengths with the Fermi Large Area
 Telescope \citep{goodenough09}. Either it is a sign for 
 the detection of dark matter (e.g. \citealt{leane19}), or it
 is due to a large collection of point sources, most likely
 millisecond pulsars. The latter trace the stellar mass in the 
 nuclear and boxy/peanut bulge of our Galaxy 
 (\citealt{ness16}) better than conventional dark 
 matter profiles \citep{bartels18}.
The possibility to detect such millisecond pulsars in the 
Milky Way with existing radiotelescopes, at Parkes and Green 
Bank, and new radiotelescopes, MeerKAT and SKA, has been 
examined in \cite{calore16}.

\subsection*{\LARGE 6.4 primoridal black holes}

In the early 1990s, efforts were made to constrain the mass 
of the dark matter by using microlensing of stars in the
Magellanic Clouds by dark objects in the Galactic halo,
as suggested by \cite{paczynski86}.
This was done by large collaborations of astronomers and
physicists: the MACHO (Massive Compact Halo Objects,
 \citealt{alcock93}), and the EROS (Exp\'erience 
pour la Recherche d'Objets Sombres, \citealt{aubourg93}) collaborations. The OGLE (Optical Gravitational Lensing Experiment, \citealt{udalski93}) project looks at microlensing
events in the direction of the Galactic Bulge. The results
from the EROS and MACHO projects were that some events were detected, but not enough to constitute the dark matter in
the galactic halo. 

Primordial black holes are considered a possibility for dark 
matter, and the search for it provide constraints for the
mass range of them. A recent review, 
\cite{carr21}, shows in their Figure 10 limits
on the mass fraction of the halo in primordial black holes
over a mass range of 10$^{-19}$ - 10$^{21}$ \msun.

\vspace{-0.4cm}
\section*{\LARGE 7. Modern rotation curves and $\LCDMbf$}

\subsection*{\LARGE 7.1 Our Galaxy}

The Gaia satellite produces accurate positions and proper motions of stars in our Galaxy, and, combined with surveys of radial velocities, allowed a new determination of the 
galactic rotation curve \citep{eilers19}. This rotation curve
is slowly declining at radii between 6.0 - 17.5 kpc, and a 
bit faster beyond that. A mass model indicated that the density
at the solar radius is 0.008 {\msun}/pc$^3$, i.e. the
same value as found by \citep{flynn06}. Dark matter is not the
dominant component of the mass in the inner parts. 
\cite{chen19} used classical Cephe\"ids to investigate the 
outer parts of the galactic disk, and demonstrate both the
warp and the flaring of the disk at these radii. The
flaring indicated by the Cephe\"ids is in good 
agreement with the flaring of the HI disk
as reported in \cite{wouterloot90}.

\subsection*{\LARGE 7.2 $\LCDMbf$: named problems}

We saw in section 5.2 how numerical simulations ruled out
neutrinos as a candidate for dark matter. The $\LCDM$-theory has
a number of sometimes hotly debated problems, which each have 
specific name: the core-cusp problem, the satellite problem, 
and the too-big-to-fail problem \citep{bullock17}. Since the theory is based on numerical simulations, it is the quality of these, and the approximations used to execute them in a reasonable amount of computer time, which determines the outcome.

\begin{figure*}
    \includegraphics[width=1.0\textwidth]{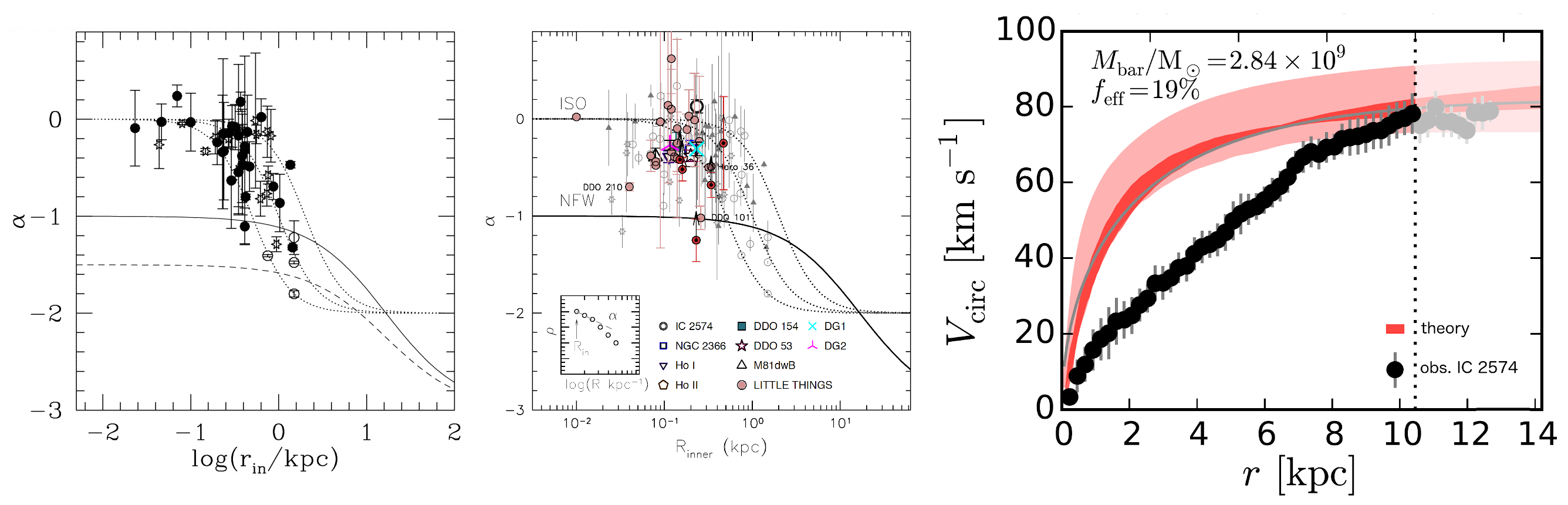}
\caption{Core-cusp comparisons for dwarf galaxies, taken from
\protect\cite{deblok01a}, at left, \protect\cite{oh15}, in the middle, and
\protect\cite{oman16}, at right. The left two panels show the values
of the inner slope of the density profile as function of the
logarithm of the radius of the innermost point on the rotation
curve. The right panel show the typical mismatch between the
rotation curve derived from a numerical simulation based on
$\LCDMbf$-theory \protect\citep{oman16}, with the observed curved for the dwarf galaxy IC 2574 from \protect\cite{oh15}.
}
\end{figure*}

In the 1990s, the cosmological numerical simulations 
produced galaxies
with dark halo profiles peaked towards the centre, which,
compared to observations, seem to pose a problem 
(\citealt{flores94, moore94}). 
It was argued in \cite{navarro97} 
that there is a universal dark matter profile
with a cusp, i.e. the central slope in log density 
should be about -1. \cite{moore98} finds a steeper slope
of -1.4. These 
results were obtained using numerical simulations with particles representing the dark matter only. However,
the result did not square with estimates made on the basis of rotation curves of late type dwarf galaxies, as reported 
e.g. in
\cite{deblok01a}, \cite{mcgaugh01}, \cite{deblok01b}, 
\cite{deblok02}, and \cite{deblok03}. These observations
clearly show that the dark matter distribution has a core,
hence the core-cusp problem.

Later work addressed the core-cusp problem in more detail,
both at the observational side, with observations at higher
angular resolution and improved sensitivity \citep{oh15}, 
and with 
numerical simulations using a hydrodynamical approach,
with the physics of the gas, the star formation and the
feedback in a so-called zoom-simulation, where the initial
conditions are taken from a prior cosmological simulation
\citep{oman16}.

The satellite problem refers to the expectation from dark
matter only simulations that there should be many low mass satellites around a massive galaxy: e.g. the number of known
satellites for our Galaxy is way too small for that.
The to-big-to-fail problem has to do with the expectation
that satellite galaxies with a virial mass of $\sim$10$^{10}$
{\msun} should have a visible baryonic component, and that
there are too few of those satellites around galaxies such
as the Milky Way. These problems have been reviewed in e.g.
\cite{bullock17}. Finally, it has been argued that a vast 
polar structure of satellite galaxies, globular clusters and 
streams around the Milky Way exists \citep{pawlowski12}. Such 
structures have also been found around the Andromeda galaxy 
and around Centaurus A \citep{pawlowski21}. These structures
are considered by some to pose a problem for the $\LCDM$
model, but the proponents of this model argue
that such structures are transient phenomena 
\citep{sawala22}. 

\vspace{-0.4cm}
\section*{\LARGE 8. Some remarks about MOND}

Part of the more recent discussions on rotation curves in spiral 
galaxies focuses on the results from a large sample of HI data for 
galaxies observed primarily with the WSRT by thesis students in 
Groningen University since the 1980s. These data, known as the 
SPARC sample (Spitzer Photometry \& Accurate Rotation Curves)
have been assembled by 
\cite{lelli16}, 
alongside near-infrared observations at {3.6\mum} with the Spitzer 
space telescope. 
\cite{mcgaugh17} 
emphasize that a plot of the observed radial acceleration
against the radial acceleration computed for the observed baryonic 
material (i.e. stellar mass density determined from the {3.6\mum} data
using a suitable mass-to-light ratio, and gas mass density determined 
from the HI distribution) is rather regular, as seen in Figure 5a.
They see this as evidence for MOND, the MOdified Newtonian Dynamics
initially developed by 
\cite{milgrom83a, milgrom83b}. 

On the other hand, people preferring to work with the ${\LCDM}$ formalism
emphasize that, in particular for late type dwarf galaxies
(e.g. \citealt{oman15}), there is
a large variety in the rotation curves, as is demonstrated by
the variation in the corresponding individual radial acceleration 
relations for each galaxy in Figure 5b.

\begin{figure}
\includegraphics[width=1.0\textwidth]{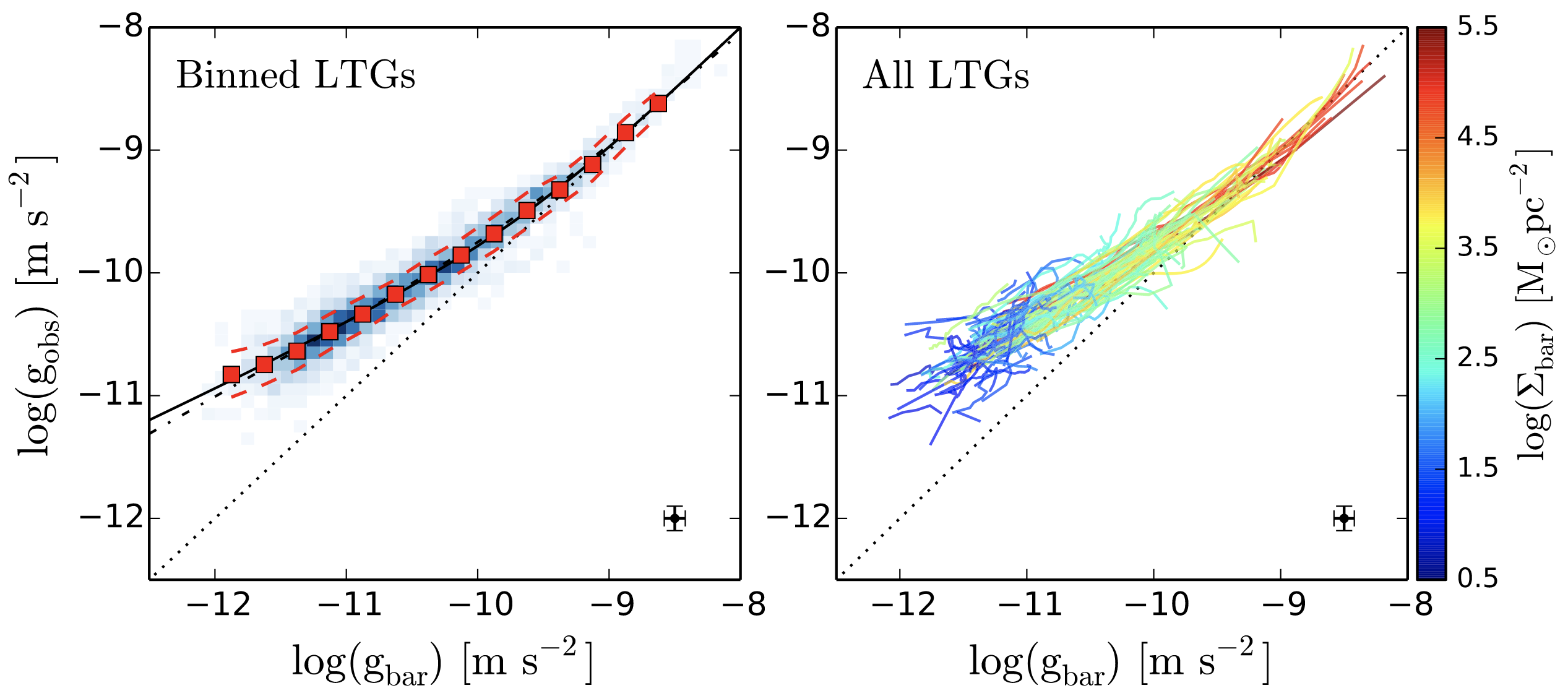}
\caption{At left, the binned radial acceleration relation using 2693 independent data points of 153 late type galaxies,
and at right, individual curves for every galaxy in the SPARC sample
(cf. \citealt{lelli17}).} 
\label{lelli-rars}
\end{figure}

Another prediction of the MOND formalism is the existence of an
external field effect (EFE). This effect due to the neighbourhood
of a galaxy affects its rotation curve, which is predicted to 
decline. 
\cite{chae20, chae21a, chae21b}  
examine this for 153 rotation curves
in the SPARC sample. They find a clear effect, and singled out two galaxies, NGC 5033 and NGC 5055, as particularly good examples \citep{chae20}.
Unfortunately, they had to revise their initial results in an erratum \citep{chae21a}.
Since I discussed both galaxies in my thesis work 
(\citealt{bosma78, bosma81a}), 
I was aware that the rotation curves of these galaxies
actually decline at larger radii, and thus rather surprised to see initially an argument that these two galaxies are in a much denser environment than average.

\cite{karachentsev21b} 
draw attention to
a particular subsample of luminous spiral galaxies with signs of declining rotation
curve, which have a radial velocity dispersion of satellites less than 55 km/s and a
poor dark matter halo. These are the galaxies
NGC 253, NGC 2683, NGC 2903, NGC 3521 and NGC 5055
(\citealt{karachentsev21a}). 
This seems in direct contradiction with the MOND EFE results
of \cite{chae21b}. 


\section*{\LARGE 9. Outlook}

\begin{figure} 
\includegraphics[width=1.0\textwidth]{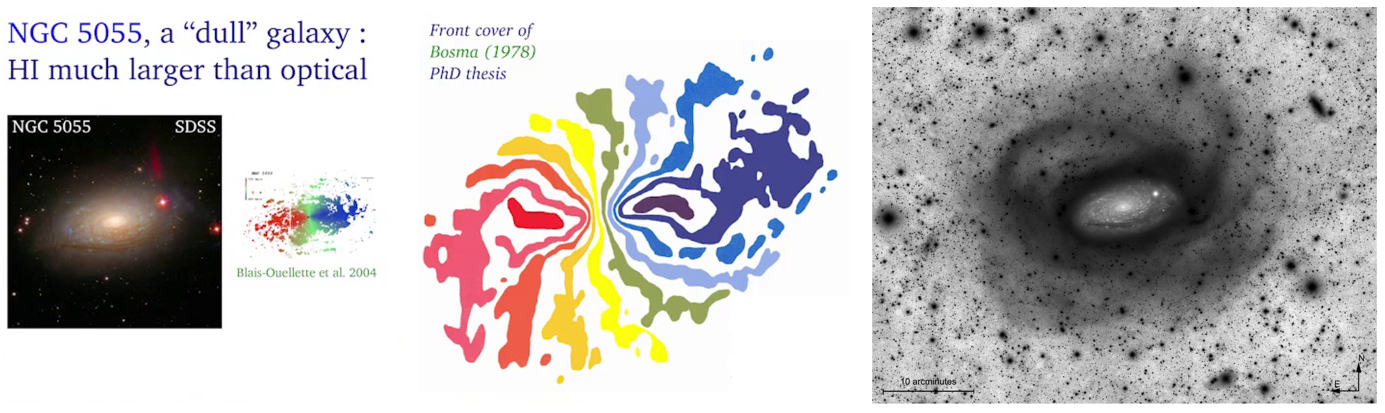}
\caption{Comparison of images of the galaxy NGC 5055, all on the same 
scale. From left to right: a multicolour optical image from SDSS, the
velocity field in H${\alpha}$ from 
\protect\cite{blaisouellete04}, 
the HI velocity field on the cover of my PhD thesis 
\protect\citep{bosma78} and a deep image in \protect\cite{karachentsev20}.
}
\label{n5055-compare}
\end{figure}

The CASTLE project (\citealt{lombardo20})
concerns the construction and
operation of a dedicated telescope for wide field observations
(2.36º x 1.56º) with a curved detector, developed at the Laboratory of Astrophysics of Marseille (LAM), which will
be installed in 2023 at the Calar Alto observatory in Spain.
The images of NGC 5055 in Figure~\ref{n5055-compare} illustrate the progress which
can be made with such a telescope in the study of the environment of
numerous nearby galaxies, highlighting the traces of interactions and
fusion of satellites with massive galaxies, as well as
exploring the low surface brightness outer parts of galaxy
disks. The striking result of the deep imaging is that the
extent of a spiral galaxy at faint levels is much larger than 
the canonical value assigned by the photometric measures 
derived half a century ago, which ended up in standard
catalogs.

\vspace{+0.50cm}

\section*{Acknowledgements}

I wish to thank the organizers of the workshop, Simon Beyne et Julien Bernard, for inviting me to review the subject of dark matter in galaxies.
I thank Lia Athanassoula for fruitful discussions about
many of the topics discussed here, and the anonymous
reviewer for a careful reading of the manuscript. This research has made use of NASA’s Astrophysics Data System Bibliographic Services.

\bibliographystyle{mnras}
\bibliography{rotcur-text-v6}

\end{document}